\documentclass[aps,twocolumn,floatfix,showpacs,amsmath,amsfonts]{revtex4}

\usepackage{graphicx}
\usepackage{bm}
\usepackage{hyperref}

\begin{document}

\title{Discovery of exceptional points in the Bose-Einstein condensation
  of gases with attractive $1/r$-interaction}

\author{Holger Cartarius}
\email{Holger.Cartarius@itp1.uni-stuttgart.de}
\author{J\"org Main}
\author{G\"unter Wunner}

\affiliation{Institut f\"ur Theoretische Physik 1, Universit\"at Stuttgart,
  70550 Stuttgart, Germany}

\date{\today}

\begin{abstract}
  The extended Gross-Pitaevskii equation for the Bose-Einstein condensation
  of gases with attractive $1/r$-interaction has a second solution which
  is born together with the ground state in a tangent bifurcation. At
  the bifurcation point both states coalesce, i.e., the energies and the
  wave functions are identical.  We investigate the bifurcation point
  in the context of exceptional points, a phenomenon known for linear
  non-Hermitian Hamiltonians.  We point out that the mean field
  energy, the chemical potential, and the wave functions show the same
  behavior as an exceptional point in a linear, non-symmetric system.
  The analysis of the analytically continued Gross-Pitaevskii equation
  reveals complex waves at negative scattering lengths below the
  tangent bifurcation.  These solutions are interpreted as a decay of
  the condensate caused by an absorbing potential.
\end{abstract}


\maketitle

\section{Introduction}
It is a well known property of Bose-Einstein condensates with 
attractive interatomic interactions that stationary solutions to the
Gross-Pitaevskii equation exist only in certain regions of the 
parameter space governing the physics of the condensates. For example,
for the case of an attractive s-wave contact interaction it was theoretically
predicted (see, e.g., \cite{Dal96,Dod96,garcia97}), and experimentally confirmed
\cite{Bra97,Sac99,Ger00}, that the condensate collapses
when, for given negative scattering length, the number of particles
becomes too large. In an alternative experiment \cite{Don01} the collapse was
induced by tuning the scattering length in the vicinity of Feshbach resonances
by adjusting an external magnetic field. Huepe et al.\ \cite{Hue99,Hue03}
have shown
that the critical parameter values where collapse occurs in fact 
correspond to bifurcation points of the solutions to the stationary
Gross-Pitaevskii equation and analyzed the linear stability of the
solutions. Quite recently it was demonstrated \cite{Pap07} that similar behavior
persists in Bose-Einstein condensates where, in addition to the 
short-range (van der Waals-like) interaction a long-range
``gravity-like'', attractive $1/r$ interaction is present. Such monopolar
quantum gases could be realized according to O'Dell et al.\ \cite{ODe00} by a
combination of 6 appropriately arranged ``triads'' of intense off-resonant
laser beams. There, too, when crossing the borderlines in the parameter
space, spanned by particle number, scattering length and trap frequency,
at which a stationary ground state of the extended Gross-Pitaevskii  equation 
comes into being, a second, excited, state of the condensate is born 
in a tangent bifurcation. 

It is the purpose of this paper to examine these
bifurcations from the point of view of the theory of ``exceptional points'',
since at the bifurcation points both the eigenvalues and the wave functions
of the two states are identical, a situation well known from studies of 
exceptional points \cite{Kato66}, which have been investigated theoretically
\cite{Ber98,Hei91,Hei99,Shu00,Hei01,Ber03,Gue07,Keck03}
and experimentally \cite{Phi00,Dem01,Dem03,Ste04,Obe96} in a wide variety
of physical systems. 

Exceptional points can appear in systems described by non-Hermitian
matrices which depend on a two-dimensional parameter space. At critical points
in the parameter space (exceptional points) a coalescence of two
eigenstates can occur, where \emph{both} the eigenvalues \emph{and} the
eigenvectors of the two states pass through a branch point singularity and
become identical. There is only one linearly independent eigenvector of
the two states at the exceptional point.

In the quantum mechanics of linear Schr\"odinger equations, exceptional
points are investigated in open systems in which resonances exist.
Examples are discussed, e.g., for complex atoms in laser fields
\cite{Lat95}, a double $\delta$ well \cite{Kor03}, the scattering of a beam
of particles by a double barrier potential \cite{Her06}, and the hydrogen
atom in static external fields \cite{Car07b}. The existence of resonances
is important for the occurrence of exceptional points in quantum systems
because the coalescence of two eigenstates is not possible in the case
of Hermitian Hamiltonians describing bound states. In the latter case, there
is always a set of orthogonal eigenstates which never become identical.
The situation is different for resonances, which can, e.g., be described by
\emph{non}-Hermitian Hamiltonians. In this case, the eigenstates have not
to be orthogonal and a coalescence can occur.

In this paper we will  reveal the existence of exceptional points
also in quantum systems described by {\em nonlinear} Schr\"odinger equations. 
As a model system we choose Bose-Einstein condensates
with attractive $1/r$-interaction and concentrate on the case of self
trapping, i.e., condensation without external trap, which is a
feature of such systems \cite{ODe00,Pap07}. It is shown that the
bifurcations of the two stationary solutions to the nonlinear Gross-Pitaevskii
equation at critical physical parameter values exhibit the typical structure
known from studies of exceptional points in linear systems.
We should point out that the effect is not restricted to condensates with
$1/r$-interaction.
Our model system has the great advantage that simple approximate
analytic solutions exist, which can be used for the investigation of the
exceptional points.
The branch point singularity structure can be seen directly from
the analytic terms we obtain, which is only possible for self-trapped
condensates \emph{without} external harmonic trap.

The Gross-Pitaevskii equation for self-trapped Bose-Einstein condensates with
monopolar $1/r$-interaction is introduced in Sec.\ \ref{sec:BEC_1_over_r}.
In Sec.\ \ref{sec:ep}, we give an overview on exceptional points in linear
systems and their properties, which can be used to identify them.
The complex vicinity of the bifurcation point is investigated in Sec.\
\ref{sec:tangent_bifurcation} to reveal the branch point singularity
structure of the bifurcation and to identify it as a ``nonlinear version''
of an exceptional point.
The physical interpretation of a complex absorbing potential in the
Gross-Pitaevskii equation at scattering lengths below the critical value
at the tangent bifurcation is discussed in Sec.\ \ref{sec:decay}.
Conclusions are drawn in Sec.\ \ref{sec:conclusion}.

\section{Bose-Einstein condensates with $1/r$-interaction}
\label{sec:BEC_1_over_r}
In this section we briefly review the equations and results for 
self-trapped Bose-Einstein condensates with attractive $1/r$-interaction which
are necessary for the subsequent analysis of exceptional points. The extended
Gross-Pitaevskii equation without external trap potential reads
\begin{multline}
  \left [ -\Delta_{\bm{r}} + N \left ( 8\pi
      \frac{a}{a_u} \left | \psi(\bm{r})\right |^2
      - 2 \int \mathrm{d}^3 \bm{r}' \frac{\left | \psi(\bm{r}')\right |^2}
      {\left | \bm{r} - \bm{r}' \right |} \right ) 
  \right ] \psi(\bm{r}) \\
  = \varepsilon \psi(\bm{r})\; , 
  \label{eq:extended_GP}
\end{multline}
where the natural ``atomic'' units introduced in \cite{Pap07}
were used. Lengths are measured in units of a ``Bohr radius'' $a_u$ and
energies in units of a ``Rydberg energy'' $E_u$, which are given by
\begin{equation*}
  a_u = \frac{\hbar^2}{m u}\; , \qquad E_u = \frac{\hbar^2}{2 m a_u^2} \; ,
\end{equation*}
respectively, where $u$ determines the strength of the atom-atom-interaction
\cite{ODe00}, and $m$ the mass of one boson. In Eq.\
\eqref{eq:extended_GP}, $\varepsilon$ is the chemical potential, $a$ the s-wave 
scattering length and $N$ the number of bosons. As was pointed out in
\cite{Pap07} with the use of the scaling property of the
system, the physics of a self-trapped condensate is only determined by one
parameter, namely $N^2 a/a_u$, which will be necessary for the identification
of the exceptional point. A further quantity necessary for our discussions
is the mean field energy of the self-trapped condensate, which reads
\begin{multline}
  E[\psi] = N \int \mathrm{d}^3 \bm{r}\: \psi^{*}(\bm{r}) \biggl ( 
    -\Delta_{\bm{r}} + 4\pi N\frac{a}{a_u} \left | \psi(\bm{r})\right |^2 \\ 
    - N \int \mathrm{d}^3 \bm{r}' \frac{\left | \psi(\bm{r}')\right |^2}
    {\left | \bm{r} - \bm{r}' \right |} \biggr ) \psi(\bm{r}) \; .
  \label{eq:mean_field_energy}
\end{multline}

The solutions we are interested in, viz. the two states emerging at
the tangent bifurcation, are radially symmetric. One of them is the ground
state. Thus we concentrate on radially symmetric solutions of
Eq.\ \eqref{eq:extended_GP}. In this case both analytic calculations  via a
variational method and numerically exact computations can be carried out.

\subsection{Variational solutions}
\label{sec:var}
\begin{figure}[t]
  \includegraphics[width = \columnwidth]{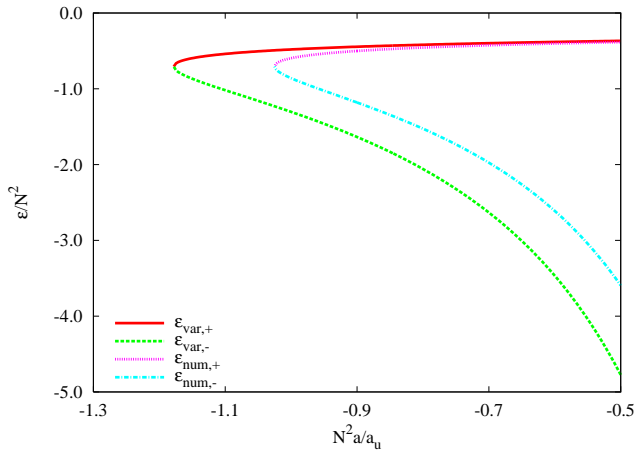}
  \caption{\label{fig:bifurcation} (Color online) Chemical potentials
    of the two solutions. The results of the variational and the numerical
    exact calculations are shown.
    Two solutions emerge in a tangent bifurcation at a critical
    value of the parameter $N^2a/a_u$. In the variational approximation
    the critical value is $N^2a/a_u = -3\pi/8$, whereas the numerical exact
    solutions emerge at $N^2a/a_u = -1.0251$ \cite{Pap07}.
    The ``atomic'' units introduced in \cite{Pap07} are used.}
\end{figure}

An analytic approximation for the two wave functions of the condensate which 
emerge at the tangent bifurcation can be obtained using a variational 
principle. Calculations with a Gaussian type orbital  were performed 
by O'Dell et al.\ \cite{ODe00} and compared with numerically accurate
solutions in \cite{Pap07}. The trial wave function is given by
\begin{equation}
  \psi(\bm{r}) = A \exp\left(-\frac{1}{2}k^2 \bm{r}^2 \right) \; ,
\end{equation} 
where
\begin{equation}
  A = \left (\frac{k}{\sqrt{\pi}}\right )^{3/2}
  \label{eq:normalization_constant_A}
\end{equation}
is the normalization constant and the variation is performed with respect
to the width represented by $k$. The two stationary values,
\begin{equation}
  \frac{E_\pm}{N^3}
   = -\frac{4}{9\pi} \frac{1\pm 2 \sqrt{1+\frac{8}{3\pi }N^2\frac{a}{a_u}}}
   {\left(1\pm\sqrt{1+\frac{8}{3\pi}N^2\frac{a}{a_u}}\right)^2} \; ,
\label{eq:GTO_mean_field_energy}
\end{equation}
of the mean field energy \eqref{eq:mean_field_energy} are obtained at
\begin{equation}
 k_\pm = \frac{1}{2} \sqrt{\frac{\pi}{2}} \frac{1}{N^2\frac{a}{a_u}}
 \left(\pm\sqrt{1+\frac{8}{3\pi}N^2\frac{a}{a_u}}-1\right) \; .
\label{eq:GTO_k}
\end{equation}
$E_+$ represents the variational approximation for the ground state of the
condensate, the second solution (excited state) is labeled $E_-$.
Furthermore, the analytical expressions for the chemical potentials of the two
solutions are given by
\begin{equation}
  \frac{\varepsilon_\pm}{N^2}
  = -\frac{4}{9\pi} \frac{5\pm 4 \sqrt{1+\frac{8}{3\pi}N^2\frac{a}{a_u}}}
   {\left(1\pm\sqrt{1+\frac{8}{3\pi}N^2\frac{a}{a_u}}\right)^2 } \; ,
\label{eq:GTO_chemical_potential} 
\end{equation}
respectively. The tangent bifurcation is obvious from the equations
\eqref{eq:GTO_mean_field_energy} and  \eqref{eq:GTO_chemical_potential}.
For negative scattering lengths with
$N^2 a/a_u < -3\pi/8$ there is no (real) result for $E$ and
$\varepsilon$, at the critical value $N^2 a/a_u = -3\pi/8$ both
solutions have the same value ($E_+ = E_-$, $\varepsilon_+ = \varepsilon_-$),
and above $-3\pi/8$, we obtain two different real solutions.
The variational chemical potential is shown as a function of the scattering
length parameter $N^2 a/a_u$ in Fig.\ \ref{fig:bifurcation}.

\subsection{Exact calculations}
\label{sec:ex}
For the search of radially symmetric wave functions, the radial part of
the integro-differential equation \eqref{eq:extended_GP} is written in the
form of two differential equations,
\begin{align}
 \Psi''(r) + \frac{2}{r}\Psi'(r)
    &= -U(r)\Psi(r) + 8\pi b|\Psi(r)|^2\Psi(r) \; ,
    \label{eq:2nd_order_ode_psi} \\
    U''(r) + \frac{2}{r}U'(r) &= -8\pi|\Psi(r)|^2\; ,
    \label{eq:2nd_order_ode_u}
\end{align}
with the parameter 
\begin{equation}
  b = \frac{a}{a_u} \; .
\end{equation}
The $N$-dependence in \eqref{eq:extended_GP} has been absorbed in the
wave function $\Psi = \sqrt{N} \psi$. 
The two second order differential equations can be transformed to first
order equations and integrated, e.g., with a Runge-Kutta-Merson algorithm
starting at, e.g., $\tilde{r}_0=10^{-19}$ slightly greater than zero with the
initial conditions
\begin{equation}
 \tilde{\Psi}(\tilde{r}_0) = 1 \; , \;
 \tilde{\Psi}'(\tilde{r}_0) = 0 \; , \;
 \tilde{U}(\tilde{r}_0) = u_0 \; , \;
 \tilde{U}'(\tilde{r}_0) = 0 \; .
\end{equation}
The value $u_0$ must be determined such that $\tilde{\Psi}(\tilde{r})$
vanishes in the limit $\tilde{r}\to\infty$.
Numerically, integration up to $\tilde{r}_{\max}\approx 15$ is sufficient.
The wave function obtained with the initial condition $\tilde{\Psi}
(\tilde{r}_0)=1$ is not normalized, which is indicated with the tilde.
Because of the scaling property \cite{Pap07} of the
problem, the normalization of a numerically obtained solution $\tilde{\Psi}$ is
possible with the help of a normalization factor $\nu$. As can be seen
with a simple calculation, the transformation
\begin{equation}
  (\Psi,r,\varepsilon,a) \to (\nu^2\Psi, \frac{r}{\nu}, \nu^2\varepsilon, 
  \frac{a}{\nu^2})
  \label{eq:scaling_invariance}
\end{equation}
leaves the system \eqref{eq:2nd_order_ode_psi}, \eqref{eq:2nd_order_ode_u}
of differential equations invariant. With the normalization condition
$||\Psi||^2 = N$ and the scaling invariance \eqref{eq:scaling_invariance},
the normalization factor is given by
\begin{equation}
 \left (\frac{\nu}{N}\right )^{-1} = ||\tilde{\Psi}||^2 = 4\pi 
 \int_0^\infty |\tilde{\Psi}(\tilde{r})|^2 \tilde{r}^2\, \mathrm{d}\tilde{r}\; 
\label{eq:normalization_integral}
\end{equation}
and the properly scaled and normalized wave function $\Psi(r)$ is obtained
as
\begin{equation}
 \Psi(r) = \nu^{2} \tilde{\Psi}(\tilde{r}/\nu)
\end{equation}
with the scaled scattering length
\begin{equation}
 N^2\frac{a}{a_u} = \frac{b}{\nu^2} \; .
\end{equation}
An explicit calculation of the integral \eqref{eq:normalization_integral} is
not required. Using the asymptotic behavior of the numerically computed
potential $\tilde{U}(\tilde{r})$ for large radial coordinates
\begin{equation}
  \tilde{U}(\tilde{r}) \approx \tilde{\varepsilon} 
  + 2\frac{N}{\nu} \frac{1}{\tilde{r}}\; ,
\label{eq:approximation_U_large_r}
\end{equation}
the factor $\nu/N$ can be calculated from
\begin{equation}
  \frac{\nu}{N} = \lim_{\tilde{r}\to\infty} \frac{-2}{\tilde{r}^2
    \tilde{U}'(\tilde{r})}\; .
\end{equation}
The correctly scaled chemical potential can also be determined with the
help of the approximation \eqref{eq:approximation_U_large_r}:
\begin{equation}
 \frac{\varepsilon}{N^2} = \left ( \frac{\nu}{N} \right )^2 
 \lim_{\tilde{r}\to\infty} \left ( \tilde{U}(\tilde{r}) + \tilde{r}
 \tilde{U}'(\tilde{r}) \right ) \; .
\end{equation}
Only for the mean field energy a further integral is required. Using the
virial theorem \cite{ODe00,Dal96} we can obtain it as
\begin{equation}
  \frac{E}{N^3} = -\frac{\langle U \rangle}{N^2} = -4\pi \left ( 
    \frac{\nu}{N}\right )^3 \int_0^\infty |\tilde{\Psi}(\tilde{r})|^2 
  \tilde{U}(\tilde{r}) \tilde{r}^2 \mathrm{d}\tilde{r}\; .
  \label{eq:mean_field_energy_integral}
\end{equation}
The numerical exact chemical potential is also shown in 
Fig.\ \ref{fig:bifurcation}.
It agrees, qualitatively, with the result of the variational approach, however,
the tangent bifurcation is shifted to a higher critical scattering length
$N^2 a/a_u=-1.0251$ \cite{Pap07}.

We will show in Sec.\ \ref{sec:tangent_bifurcation} that in both  the
variational approach and the exact calculations the tangent bifurcation at
the critical scattering length is a branch point singularity of the wave
functions, i.e., an  {\em exceptional point}.

\section{Exceptional points}
\label{sec:ep}
\begin{figure}[t]
  \includegraphics[width = \columnwidth]{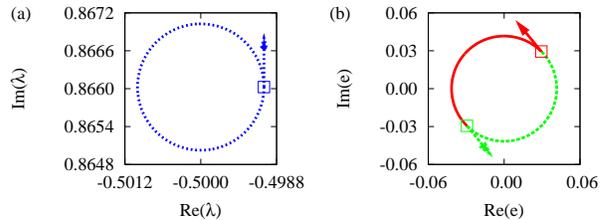}
  \caption{\label{fig:2d_circle} (Color online) Eigenvalues of the
    two-dimensional model defined in Eq.\ \eqref{eq:2d_model} with $c=1$ 
    for a circle around the exceptional point. (a) Circle in the parameter
    space with radius $\varrho = 10^{-3}$. The starting point $\lambda(0)$
    is marked with an open square and the direction of progression is indicated 
    with the arrow.
    (b) Path of the eigenvalues $e_{1,2}$ for the parameter values
    on the circle. The eigenvalues which belong to the first parameter
    value $\lambda(0)$ are labeled with open squares and the
    arrows point in the direction of progression.}
\end{figure}

To make our presentation self-contained we recapitulate the 
essential concepts on exceptional points and  their properties known from
linear systems, which will be important for
the comparison with the coalescence of two quantum states in a nonlinear
Schr\"odinger equation in Sec.\ \ref{sec:tangent_bifurcation}. Usually complex
symmetric matrices or complex symmetric matrix representations of Hamiltonians
are used to investigate exceptional points in quantum systems (see, e.g.,
\cite{Hei99,Kor03,Car07b}).
Here, we treat the more general case of non-symmetric matrices, considered
before \cite{Hei04,Sey05,Mail05} also for systems with time reversal
symmetry breaking \cite{Har04,Hei06,Ber06},
and compare the phase behavior of the eigenvectors with the symmetric
case. It is instructive to discuss a simple
two-dimensional model which exhibits exceptional points. The example
is provided by the $2\times 2$-matrix
\begin{equation}
  \bm{M}(\lambda) = \begin{pmatrix} 1 & \lambda \\
    c + \lambda & -1 \end{pmatrix}.
  \label{eq:2d_model}
\end{equation}
In \eqref{eq:2d_model}, $\lambda$ is a  complex parameter, and  the complex
number $c$ is introduced to distinguish between the general non-symmetric
case ($c \ne 0$) and a symmetric matrix ($c = 0$) in the discussion below.
The eigenvalues read
\begin{equation}
    e_{1,2}(\lambda) =  \pm \sqrt{1 + c\lambda + \lambda^2}
\end{equation}
and are obviously two branches of one analytic function. Corresponding
non-normalized eigenvectors of the two eigenvalues are given by
\begin{equation}
  \bm{x}_{1,2}(\lambda) = \left ( \begin{array}{c} 
      -\lambda \\ 1\mp \sqrt{1 + c\lambda + \lambda^2} \end{array} \right )
  \; .
  \label{eq:eigenvectors_model}
\end{equation}
There are two exceptional points for $\lambda_{A,B} = -c/2 \pm 
\sqrt{c^2/4-1}$. At these parameter values, both the eigenvalues
$e_{1,2}$ and the eigenvectors $\bm{x}_{1,2}$ pass through a branch
point singularity. For the further discussions we concentrate on the case
$\lambda = \lambda_A$. Then the degenerate eigenvalues have the
value $e_{1,2}(\lambda_A) = 0$ and the two eigenvectors
$\bm{x}_{1,2}$ are
\begin{equation}
  \bm{x}_{1,2}(\lambda_A) = \begin{pmatrix} c/2 - \sqrt{c^2/4-1} \\ 
      1 \end{pmatrix}  \; .
\end{equation}
An important property of exceptional points, which follows from the branch
point singularity structure, is the permutation of the two eigenvalues if
the exceptional point is encircled in the parameter space \cite{Kato66}.
To illustrate this effect we follow the paths of the eigenvalues
$e_{1,2}$ in the complex plane for a circle with radius $\varrho$
around the critical value $\lambda_A$, which can be represented by
\begin{equation}
  \lambda_\varrho(\varphi) = -c/2 + \sqrt{c^2/4-1}
  + \varrho\, \mathrm{e}^{\mathrm{i} \varphi}\; .
  \label{eq:parameter_space_circle_2d}
\end{equation}
An approximation for small radii $\varrho \ll | 2\sqrt{c^2/4-1} |$
leads to
\begin{equation}
  \begin{aligned}
    e_1(\lambda_\varrho(\varphi)) &= \sqrt{2 \varrho \sqrt{c^2/4-1}}\,
    \mathrm{e}^{\mathrm{i} (\varphi/2+\pi/4)} \; ,\\
    e_2(\lambda_\varrho(\varphi)) &= \sqrt{2 \varrho \sqrt{c^2/4-1}}\,
    \mathrm{e}^{\mathrm{i} (\varphi/2+5\pi/4)} \; .
  \end{aligned}
  \label{eq:eigenvalues_small_r}
\end{equation}
If a full circle in the complex parameter space $\lambda$ is traversed, the
paths of the eigenvalues form a semicircle as can be seen from Eq.\
\eqref{eq:eigenvalues_small_r}.
Neither of the two eigenvalues passes through a closed loop. The
first arrives, after the parameter space loop, at the starting point of the
second one and vice versa. Two circles in the $\lambda$-space
($\varphi = 0\dots 4\pi$) are required to obtain a full circle of \emph{one}
of the eigenvalues. The situation is shown in Fig.\ \ref{fig:2d_circle} for
$c = 1$.
In the parameter space the circle \eqref{eq:parameter_space_circle_2d}
is traversed once with a radius $\varrho = 10^{-3}$, as 
 plotted in Fig.\ \ref{fig:2d_circle}~(a).
The paths of the two eigenvalues in the complex $e$-space is
shown in Fig.\ \ref{fig:2d_circle}~(b). The semicircle structure expected
from approximation \eqref{eq:eigenvalues_small_r} is clearly visible. 

For exceptional points in linear systems there is a noteworthy
difference for  complex symmetric and non-symmetric matrices.
In the first case, which is obtained for $c = 0$ in the model
\eqref{eq:2d_model} and which was realized in the resonances investigated
in microwave cavities \cite{Dem01} and in the hydrogen atom in crossed
electric and magnetic fields \cite{Car07b}, there is always a distinct
phase behavior of the eigenvectors. If an exceptional point is
encircled, the two eigenvectors are interchanged (as expected from the
behavior of the energy eigenvalues) and, additionally, \emph{one} of the
two eigenvectors changes its sign. This effect is demonstrated in
\cite{Hei99} and can be summarized with, e.g., 
\begin{equation}
  \left [ \bm{x}_1, \bm{x}_2 \right ]  
  \qquad \overset{\text{circle}}{\longrightarrow} \qquad
  \left [ -\bm{x}_2, \bm{x}_1 \right ]\; .
  \label{eq:sign_change}
\end{equation}
Four circles around the exceptional point in the parameter space
are needed to restore the original situation $\left [ \bm{x}_1, \bm{x}_2
\right ]$. A direct verification of the effect was performed with 
microwave cavity experiments \cite{Dem01}, in which a visualization of the
wave functions was possible. The phase behavior can be obtained
for the vectors $\bm{x}_i$ normalized with the Euclidean norm without
complex conjugation $N_i = \sqrt{\bm{x}_i^T\bm{x}_i}$, where $\bm{x}_i^T$ is
the transpose of the vector $\bm{x}_i$ (cf. \cite{Keck03}). This normalization
fixes the phase and is the method required for the comparison with the
nonlinear system studied in this paper (cf. Sec.\ \ref{sec:tangent_bifurcation}
and appendix).

For the non-symmetric case the change in sign does not appear
for a vector whose phase has been fixed with the same method, i.e., the
Euclidean norm without complex conjugation.
A non-symmetric matrix is obtained in the model \eqref{eq:2d_model} for
$c \neq 0$. Then, there is only a permutation of the eigenvectors, which
can be summarized with
\begin{equation}
  \left [ \bm{x}_1, \bm{x}_2 \right ]  
  \qquad \overset{\text{circle}}{\longrightarrow} \qquad
  \left [ \bm{x}_2, \bm{x}_1 \right ]\; .
  \label{eq:no_sign_change}
\end{equation}
This is shown analytically for a small circle in the appendix.
Note that the phase behavior \eqref{eq:no_sign_change} does not represent
the ``geometric phase'' of an eigenvector during the circle around
an exceptional point, which has been calculated in Ref.\ \cite{Mail05}.

\begin{figure}[t]
  \includegraphics[width = \columnwidth]{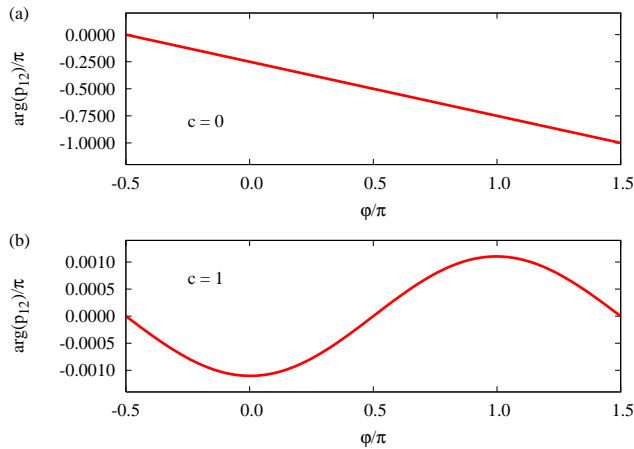}
  \caption{\label{fig:phases_model} (Color online) Phase of the product
    $p_{12}$ for a circle in the parameter space of type
    \eqref{eq:parameter_space_circle_2d} with radius $\varrho = 10^{-3}$
    around the exceptional point. (a) Symmetric matrix with $c=0$. The
    phase changes by $\pi$ indicating the change in sign of one of the
    eigenvectors from \eqref{eq:sign_change}. (b) Non-symmetric
    matrix with $c=1$. There is no change in sign as noted in 
    \eqref{eq:no_sign_change}.}
\end{figure}

A demonstration of the phase behavior is possible with the product
\begin{equation}
  p_{12} = \bm{x}_1^T \begin{pmatrix} 0 & 1 \\ 1 & 0 \end{pmatrix}
  \bm{x}_2\; .
  \label{eq:product_model}
\end{equation}
The non-diagonal matrix ensures, that the product $p_{12}$ does not vanish in
the symmetric case. If the change in sign is present, the phase of $p_{12}$
changes its value by $\pi$ when  a full circle around the exceptional point
is traversed.
In Fig.\ \ref{fig:phases_model} the behavior of the product is shown for the
example \eqref{eq:2d_model}. The change in sign is obvious in the symmetric
case $c = 0$ and does not appear for the non-symmetric choice $c=1$.

\section{Analytic continuation of the Gross-Pitaevskii system}
\label{sec:tangent_bifurcation}
In the extended Gross-Pitaevskii equation \eqref{eq:extended_GP}, both the
variational and the numerical calculations suggest the existence of an
exceptional point at the tangent bifurcation. For the critical parameter
value, we obtain two identical states. The chemical potentials, the mean
field energies and the wave functions are identical. In contrast to
exceptional points in linear systems described in Sec.\ \ref{sec:ep} the
coalescence of the two states of the nonlinear Schr\"odinger equation
\eqref{eq:extended_GP} can appear for a purely real wave function at a real
energy in a one-dimensional parameter space. This is a consequence of the
nonlinearity of the Schr\"odinger equation.

If we want to check whether or not the degeneracy in the nonlinear system has
the same branch point singularity structure as exceptional points in open
quantum systems described by a linear Schr\"odinger equation, we have
to extend the parameter $N^2 a/a_u$ to complex values and to investigate
the complex vicinity of the degeneracy.

The analytic continuation of the Gross-Pitaevskii system is a nontrivial task.
The Gross-Pitaevskii equation \eqref{eq:extended_GP} contains the square
modulus of the wave function $\psi$ and is therefore a non-analytic function
of $\psi$. It has been argued that the tempting simple replacement of $|\psi|^2$
with $\psi^2$ is valid only with the assumption that the entire
wave function is real valued \cite{Moi05,Schla06}.
Here, we suggest the following procedure for complex wave functions.

Any complex wave function can be written as
\begin{equation}
 \psi(\bm{r}) = \mathrm{e}^{\alpha(\bm{r})+\mathrm{i}\beta(\bm{r})} \; ,
\label{eq:psi_cmplx}
\end{equation}
where the real functions $\alpha(\bm{r})$ and $\beta(\bm{r})$ determine the
amplitude and phase of the wave function, respectively.
The complex conjugate and the square modulus of $\psi(\bm{r})$ read
\begin{equation}
 \psi^\ast(\bm{r}) = \mathrm{e}^{\alpha(\bm{r})-\mathrm{i}\beta(\bm{r})} \; , \;
 |\psi(\bm{r})|^2 = \mathrm{e}^{2\alpha(\bm{r})} \; .
\label{eq:psi_mod2}
\end{equation}
Using the ansatz \eqref{eq:psi_cmplx} the Gross-Pitaevskii system can be
written as two coupled nonlinear differential equation for $\alpha(\bm{r})$ and
$\beta(\bm{r})$, however, without any complex conjugate or square modulus.
These equations can now be  continued analytically by allowing for
complex valued functions $\alpha(\bm{r})$ and $\beta(\bm{r})$.
This implies that Eq.\ \eqref{eq:psi_mod2} {\em without} complex conjugate 
of $\alpha(\bm{r})$ and $\beta(\bm{r})$ is formally used for the calculation of
$\psi^\ast$ and $|\psi(\bm{r})|^2$, and thus the square modulus of $\psi$
can become complex.
The physical interpretation of a complex absorbing potential in the
Gross-Pitaevskii system will be discussed below in Sec.\ \ref{sec:decay}.

It should be noted that the above procedure is analogous to the complex
scaling method, viz.\ the replacement $r\to r\mathrm{e}^{\mathrm{i}\theta}$ in
the Hamiltonian, used for the calculation of resonances in open quantum
systems \cite{Rei82,Moi98}.
For states on the left hand side of operators there is no complex
conjugation of the $r\mathrm{e}^{\mathrm{i}\theta}$ arising from the scale
transformation but complex conjugation is applied to the intrinsically
complex part of the function.

In the following we will encircle the branch point singularity at the
tangent bifurcation.
For the variational approach discussed in Sec.\ \ref{sec:var} the 
Eqs.\ \eqref{eq:GTO_mean_field_energy}, \eqref{eq:GTO_k}, 
and \eqref{eq:GTO_chemical_potential}
are straightforwardly extended to complex values of the parameter
$N^2 a/a_u$, yielding complex results for the chemical potential,
the mean field energy, and the wave functions.
The situation is more complicated for the exact calculations
in Sec.\ \ref{sec:ex} because the differential equations 
\eqref{eq:2nd_order_ode_psi} and \eqref{eq:2nd_order_ode_u}
do not directly depend on the scaled scattering length $N^2 a/a_u$
but on the parameter $b$.
For complex $b$ the differential equations 
\eqref{eq:2nd_order_ode_psi} and \eqref{eq:2nd_order_ode_u}
can be  continued analytically resulting in complex wave functions.
Then, the two complex parameters $u_0$ and $b$ are determined in a 
multidimensional root search problem, to fulfill the condition of
vanishing wave function in the limit $r\to\infty$, and to achieve 
the given value of $N^2 a/a_u$.

\subsection{Branch point singularity structure of the energies}
\begin{figure}[t]
  \includegraphics[width = \columnwidth]{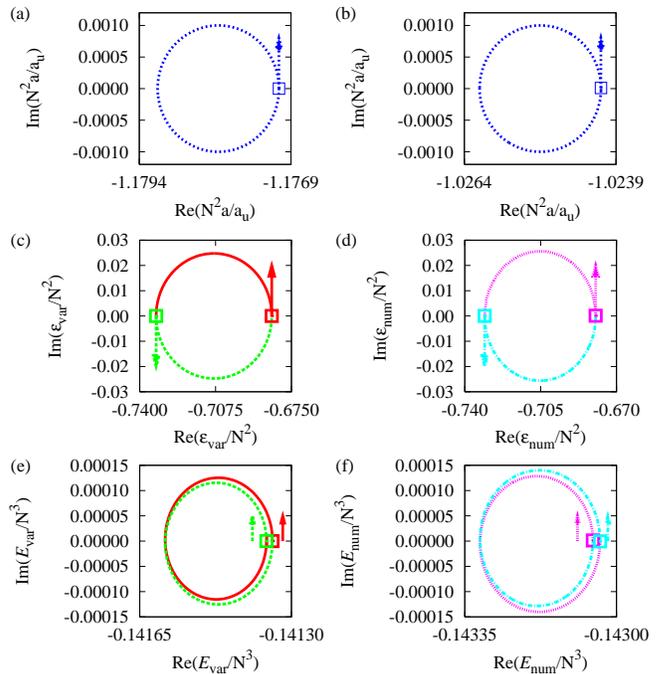}
  \caption{\label{fig:circles}(Color online) Chemical potential and
    mean field energy for a circle in the complex $N^2 a/a_u$ parameter
    space. The first row shows the parameter space circle of type 
    \eqref{eq:parameter_space_circle} with $\varrho = 10^{-3}$ used for the
    variational (a) and the numerical exact (b) calculations. The
    starting point is marked with an open square and the direction of
    progression is indicated with an arrow. In the second row, the chemical
    potential $\varepsilon/N^2$ for the variational (c) and numerical (d)
    solutions is plotted. The variational (e) and numerical (f)
    mean field energies $E/N^3$ are presented in the third row. Each of
    the two solutions is drawn with a different line. The results obtained
    for the first parameter value on the circle are marked with an open
    square and the arrows point in the direction of progression. A
    permutation of the two solutions is present for the mean field energy
    as well as for the chemical potential indicating the existence of an
    exceptional point. All energies are given in units of the ``Rydberg
    energy'' $E_u$ mentioned in Sec.\ \ref{sec:BEC_1_over_r}.}
\end{figure}

The variational solutions \eqref{eq:GTO_mean_field_energy} for the mean field
energy, and \eqref{eq:GTO_chemical_potential} for the chemical potential
show a branch point singularity at the bifurcation point. For a circle around
the critical value of the complex parameter $N^2a/a_u$ the typical 
permutation of the two solutions is expected and found, as is shown
in Fig.\ \ref{fig:circles} for the variational approximation as well as for
the numerical exact calculations.
The circle in the parameter space is defined by 
\begin{equation}
  N^2a/a_u = (N^2a/a_u)_\mathrm{c} + \varrho\, \mathrm{e}^{i\varphi}\; ,\qquad 
  \varphi = 0\dots 2\pi \; ,
  \label{eq:parameter_space_circle}
\end{equation}
where $(N^2a/a_u)_\mathrm{c}$ is set to the critical value
$(N^2a/a_u)_\mathrm{var} = -3\pi/8 = -1.1780$ in the variational and to
$(N^2a/a_u)_\mathrm{num} = -1.0251$ in the numerical exact calculations,
respectively. We used the radius $\varrho = 10^{-3}$. 
Both cases are shown in Fig.\ \ref{fig:circles} (a) and (b).
In Fig.\ \ref{fig:circles} (c) and (d) the variational and numerical
solutions for the chemical potential are shown.
The results were obtained for different parameter values located on the
circle. As can be seen in the figure, the permutation of the two values of
the chemical potential appears very clearly. Each of the two solutions
$\varepsilon_+$ and $\varepsilon_-$ traverses a path in the complex energy
plane similar to a semicircle. A fractional power series expansion of the
variational result \eqref{eq:GTO_chemical_potential} for small radii,
\begin{multline}
  \frac{\varepsilon_\pm}{N^2} = -\frac{20}{9\pi} \pm \frac{8}{3\pi} \cdot
  \sqrt{\varrho}\, \mathrm{e}^{\mathrm{i}\varphi/2} - \left ( \frac{4}{3\pi}
    + \frac{128}{27\pi^2} 
  \right ) \cdot \sqrt{\varrho}^2\, \mathrm{e}^{\mathrm{i} \varphi} \\
  \pm \left ( \frac{8}{9\pi} - \frac{64}{9\pi^2} \right ) \cdot 
  \sqrt{\varrho}^3\, \mathrm{e}^{(3/2)\mathrm{i} \varphi} + \mathrm{O}
  \left (\sqrt{\varrho}^4 \right )\, ,
\end{multline}
confirms this finding.  The term $\frac{8}{3\pi} \cdot \sqrt{\varrho}\,
\mathrm{e}^{i\varphi/2}$, which dominates the path of the eigenvalue for a
value $\varrho \ll 1$, leads to a semicircle. 

For the mean field energy the permutation is also present but the structure
of the paths of the two solutions is different. As can be seen from the
fractional power series expansion of the analytic solution 
\eqref{eq:GTO_mean_field_energy},
\begin{multline}
  \frac{E_\pm}{N^3} = -\frac{4}{9\pi} + 0 \cdot 
  \sqrt{\varrho}\, \mathrm{e}^{\mathrm{i}\varphi/2}
  + \frac{32}{27\pi^2} \cdot \sqrt{\varrho}^2\, \mathrm{e}^{\mathrm{i} \varphi} \\
  \pm \left ( \frac{4}{9\pi} - \frac{32}{9\pi^2} \right ) \cdot 
  \sqrt{\varrho}^3\, \mathrm{e}^{(3/2)\mathrm{i} \varphi} 
  + \mathrm{O}\left (\sqrt{\varrho}^4 \right )\, ,
\end{multline}
the first order term with the phase term $\mathrm{e}^{i\varphi/2}$ vanishes.
The lowest non-vanishing order has the phase factor 
$\mathrm{e}^{\mathrm{i} \varphi}$, which leads to a closed curve for
a complete circle in the parameter space ($\varphi = 0\dots 2\pi$). The
third order term is the lowest order responsible for a permutation.
As a consequence, it becomes more and more difficult to see a permutation for
decreasing radii $\varrho$. Nevertheless, the permutation of the two values is
present. The same result can be seen in Fig.\ \ref{fig:circles} (e) and (f).
The dominating structure is given by the circle following from the second
order term but the permutation is clearly visible even for the small radius
$\varrho = 10^{-3}$.

The exceptional point discussed here was discovered for a real 
value of the parameter $N^2 a/a_u$, however, it is still an isolated
point in the two-dimensional parameter space spanned by the real and
imaginary parts of $N^2 a/a_u$. This finding is in full agreement with
non-Hermitian linear systems, in which the co-dimension of exceptional
points is two (see, e.g., \cite{Sey05}). The analytic continuation is
required to confirm the branch point singularity structure. As mentioned
above, the appearance on the real axis is possible because of the
nonlinearity of the Schr\"odinger equation.

\subsection{\label{sec:eigenvectors}Behavior of the wave functions}
The eigenfunctions of the extended stationary Gross-Pitaevskii equation
\eqref{eq:extended_GP} are not orthogonal. They behave more like the
eigenstates of a non-symmetric linear system and the phase
behavior \eqref{eq:no_sign_change} without a change in sign is expected. A
possibility to check this is to calculate the complex overlap 
integral for the two non-orthogonal normalized states
\begin{equation}
  O_{12} = 4\pi \int_{0}^{\infty}
  \psi_1(r) \psi_2(r)\, r^2 \mathrm{d}r
  \label{eq:overlap_integral}
\end{equation}
for parameter values located on a circle of the type
\eqref{eq:parameter_space_circle_2d}. 
For this calculation it is important to note that the numerical
wave functions calculated with the method described above have a phase,
which is fixed by the normalization factor $\nu$ determined by Eq.\
\eqref{eq:normalization_integral} without complex conjugation for the
analytic continuation of the extended Gross-Pitaevskii model (cf. Eq.\ 
\eqref{eq:psi_mod2}). A  similar procedure is used for the variational result,
where the normalization constant $A$ in \eqref{eq:normalization_constant_A}
is calculated with a normalization integral without complex conjugation.

If one of the two states
changes its sign during the traversal of the loop, the phase of the 
complex value $O_{12}$ changes its value by $\pi$ similarly to the
phase behavior of the product $p_{12}$ from Eq.\ \eqref{eq:product_model}.
If this is not the case, the phase returns to
its original value at the end of the loop. 

\begin{figure}[t]
  \includegraphics[width = \columnwidth]{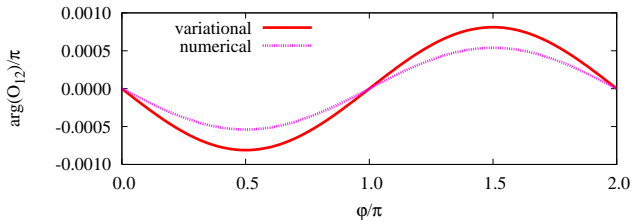}
  \caption{\label{fig:phases}(Color online) Phase $\arg(O_{12})$ of the
    overlap integral $O_{12}$ defined in Eq.\ \eqref{eq:overlap_integral}.
    The path in the parameter space $N^2 a/a_u$ is a circle
    with the critical value as center point and $\varrho = 10^{-3}$. The
    angle on the circle is denoted by $\varphi$. Both the variational 
    and the numerical results show that the phase of $O_{12}$ returns
     to its initial value at the end of the circle and that the change in
    sign from \eqref{eq:sign_change} is not present as expected (see text).}
\end{figure}

In Fig.\ \ref{fig:phases} the phase of the integral
\eqref{eq:overlap_integral} is plotted for the circle defined in
Eq.\ \eqref{eq:parameter_space_circle} with radius $\varrho = 10^{-3}$.
The result shows clearly that the phase of $O_{12}$ returns to its initial
value after one circle around the exceptional point, demonstrating that 
no change in sign of the eigenvectors occurs. Thus, the wave
functions behave in the same way as for exceptional points in linear systems
described by a non-symmetric matrix (see Sec.\ \ref{sec:ep}).

\section{Decay of the condensate}
\label{sec:decay}
The Gross-Pitaevskii equation \eqref{eq:extended_GP} depends on the
scattering length $N^2 a/a_u$ of the contact potential.
In Sec.\ \ref{sec:tangent_bifurcation} that parameter has been extended
to complex values, although the physical scattering length is always real.
There is, however, a physical situation where complex continuation is
needed for real $N^2 a/a_u$, viz.\ when the scattering length is below
the critical value at the tangent bifurcation.
For $N^2 a/a_u<-3\pi/8=-1.1780$ the variational solutions
\eqref{eq:GTO_mean_field_energy} and \eqref{eq:GTO_chemical_potential}
for the mean field energy and the chemical potential become complex.
Those parameters become complex also for the analytically continued
exact calculations at $N^2 a/a_u<-1.0251$. A complex chemical potential
as a signature of a decaying condensate has already been discussed 
in Ref.\ \cite{Cra05}.

\begin{figure}[t]
  \includegraphics[width = \columnwidth]{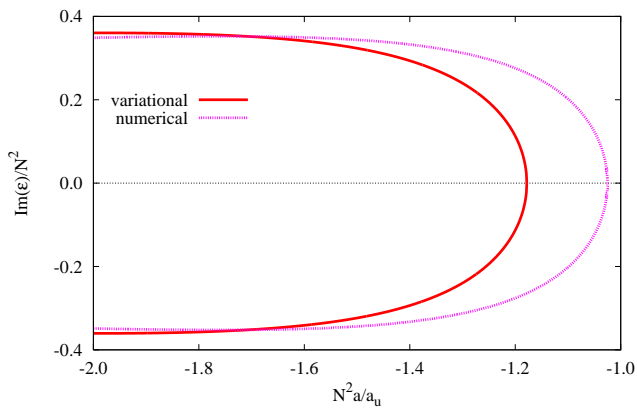}
  \caption{\label{fig:decay}(Color online) Imaginary part of the chemical
  potential for real $N^2 a/a_u$ below the bifurcation point obtained with the
  variational and the numerical calculations. Both the continuation of the
  ground state and the continuation of the excited state are plotted. One
  is the complex conjugate of the other.}
\end{figure}

The standard physical interpretation of complex eigenenergies is that
they describe decaying resonances in open systems or systems with
absorbing potentials.
The imaginary part of the energy $E$ is related to the width $\Gamma$,
decay rate $\lambda$, and lifetime $T$ of the resonance by
$\Gamma=\hbar\lambda=\hbar/T=-2\,\mathrm{Im}\;E$.
Both the real and complex wave function of the Bose-Einstein condensate
above and below the tangent bifurcation vanishes at large radius $r$,
i.e., the wave function cannot describe a decay of the condensate by
outgoing particles.
How is a decay possible?
The solution is that for a complex wave function $\psi(\bm{r})$, using
the rules \eqref{eq:psi_mod2} for complex conjugate and square modulus,
the effective potential
\begin{equation}
 V_\mathrm{eff}(\bm{r}) = 8\pi N\frac{a}{a_u} |\psi(\bm{r})|^2
  - 2N\int \mathrm{d}^3\bm{r}'\,\frac{|\psi(\bm{r}')|^2}{|\bm{r}-\bm{r}'|}
\end{equation}
becomes an {\em absorbing} potential.
Thus, the decay is an internal collapse of the condensate.
The physical interpretation might be that at large negative scattering
lengths the contact potential is so attractive that the atoms of the
condensate form molecules or clusters
as discussed, e.g., in Refs.\ \cite{Don01,Cla02,Cra05}.
The decreasing number of atoms means the decay of the condensate.
The imaginary part of the chemical potential as a function of the scattering
length is presented in Fig.\ \ref{fig:decay} for the variational as well as 
for the exact calculation. As was the case for parameter values above the
critical point, there are two solutions. Below the critical $N^2 a/a_u$ one
is the complex conjugate of the other. 

\section{Conclusion}
\label{sec:conclusion}
We have investigated the complex vicinity of the tangent bifurcation appearing
in the extended Gross-Pitaevskii equation for self-trapped Bose-Einstein
condensates with attractive $1/r$ interaction. The chemical potential
and the mean field energy show the typical behavior of energies at
an exceptional point. Both quantities pass through a branch point singularity
at the critical value of the system's parameter $N^2 a/a_u$. An analysis of the
phase behavior of the corresponding wave functions showed that the
bifurcation point has the same properties as an exceptional point of
a non-Hermitian non-symmetric matrix. Thus, we conclude that we have
identified the bifurcation point as a ``nonlinear version'' of
an exceptional point.

For the investigation of the bifurcation point an analytic continuation
of the Gross-Pitaevskii system was used. As we have discussed, an 
analytic continuation of the wave functions is not only required for
complex values of $N^2 a/a_u$ but is necessary for real parameter
values below the bifurcation point. Then, a complex absorbing effective
potential is obtained indicating an internal collapse of the potential. An
investigation of the dynamics of the system similar to Refs.\
\cite{Hue99,Hue03} will give more insight in the stability of the states and
is the topic of current studies.

In this paper we have identified the bifurcation points appearing in
self-trapped monopolar Bose-Einstein condensates as exceptional points. 
There is good reason to believe that quite generally the critical
parameter values of attractive Bose-Einstein condensates where collapse
of the condensates sets in are associated with exceptional points. 
The existence of a bifurcation point is known for condensates
in a harmonic trap without $1/r$-interaction \cite{Hue99,garcia97}
and for condensates with dipole-dipole interaction and harmonic trap.
We have checked that the bifurcation points appearing in
variational approximations of these systems are exceptional points.
Clearly this should be confirmed by detailed analysis including
numerically exact calculations for attractive Bose-Einstein condensates
based on the techniques applied in this paper.

\begin{acknowledgments}
  This work was supported by Deutsche Forschungsgemeinschaft. H.C. 
  is grateful for support from the Landesgraduiertenf\"orderung of
  the Land Baden-W\"urttemberg.
\end{acknowledgments}

\appendix*

\section{Phase of the eigenvectors on the circle}

To be able to compare the phase behavior of the wave functions of the
nonlinear Gross-Pitaevskii system with the eigenvectors of a matrix, we 
fix the phase of the vectors with the same method as is discussed in
Sec.\ \ref{sec:tangent_bifurcation} for the wave functions of the analytic
continuation of the extended Gross-Pitaevskii model. This is done with
the Euclidean norm without complex conjugation
$N_i = \sqrt{\bm{x}_i^T\bm{x}_i}$ of the eigenvectors, which is the
adequate counterpart of the normalization factors $\nu$ from Eq.\
\eqref{eq:normalization_integral} and $A$ from Eq.\
\eqref{eq:normalization_constant_A}, where (formally) the complex conjugation
is ignored. Then, the normalized eigenvectors have the form 
\begin{align*}
    \bm{x}_{1,2}(\lambda) = \frac{1}{N_{1,2}} \left ( \begin{array}{c} 
        -\lambda \\ 1\mp \sqrt{1 + c\lambda + \lambda^2} \end{array} \right )
    \intertext{with}
    N_{1,2} = \sqrt{\lambda^2 + \left ( 1\mp \sqrt{1+c\lambda+\lambda^2} 
      \right )^2 }
\end{align*}
For a small circle around the exceptional point $\lambda_A = -c/2 + 
\sqrt{c^2/4-1}$ we calculate the fractional power series expansion in
$\varrho^{1/4}$ of the normalized vectors and obtain in the non-symmetric
case ($c\neq 0$)
\begin{align*}
  \bm{x}_1(\varphi) &= \begin{pmatrix}
    \frac{1}{c}\sqrt{\eta/2} + \frac{2}{c}\sqrt{\kappa/\eta}
    \sqrt{\varrho} e^{\mathrm{i}\varphi/2} 
    + \mathrm{O}(\varrho^{3/4}) \\
    \sqrt{2/\eta} - \frac{1}{c^2} \sqrt{\kappa \eta}
    \sqrt{\varrho} e^{\mathrm{i}\varphi/2} 
    + \mathrm{O}(\varrho^{3/4})
  \end{pmatrix} \; , \\
  \bm{x}_2(\varphi) &= \begin{pmatrix}
    \frac{1}{c}\sqrt{\eta/2} - \frac{2}{c}\sqrt{\kappa/\eta} 
    \sqrt{\varrho} e^{\mathrm{i}\varphi/2} 
    + \mathrm{O}(\varrho^{3/4}) \\
    \sqrt{2/\eta} + \frac{1}{c^2}\sqrt{\kappa \eta}
    \sqrt{\varrho} e^{\mathrm{i}\varphi/2}
    + \mathrm{O}(\varrho^{3/4})
  \end{pmatrix} 
\end{align*}
with $\kappa = \sqrt{c^2/4-1}$, and $\eta = c^2-2c\kappa$.

The traversal of a single circle ($\varphi = 0\dots 2\pi$) obviously leads
to a permutation of the eigenvectors as summarized in \eqref{eq:sign_change}. 

In the symmetric case ($c=0$), a fractional power series expansion of the
two normalized eigenvectors looks different. The result is 
\begin{align*}
  \bm{x}_1(\varphi) &= \begin{pmatrix}
    \frac{e^{\mathrm{i}7\pi/8}}{2^{3/4}}\frac{e^{-\mathrm{i}\varphi/4}}{\varrho^{1/4}}
    + \frac{e^{\mathrm{i}9\pi/8}}{2^{5/4}} \varrho^{1/4}
    e^{\mathrm{i}\varphi/4} + \mathrm{O}(\varrho^{3/4}) \\
    \frac{e^{\mathrm{i}11\pi/8}}{2^{3/4}}\frac{e^{-\mathrm{i}\varphi/4}}{\varrho^{1/4}}
    + \frac{e^{\mathrm{i}5\pi/8}}{2^{5/4}} \varrho^{1/4}
    e^{\mathrm{i}\varphi/4} + \mathrm{O}(\varrho^{3/4})
  \end{pmatrix} \; , \\
  \bm{x}_2(\varphi) &= \begin{pmatrix}
    \frac{e^{\mathrm{i}11\pi/8}}{2^{3/4}}\frac{e^{-\mathrm{i}\varphi/4}}{\varrho^{1/4}}
    + \frac{e^{\mathrm{i}5\pi/8}}{2^{5/4}} \varrho^{1/4}
    e^{\mathrm{i}\varphi/4} + \mathrm{O}(\varrho^{3/4}) \\
    \frac{e^{\mathrm{i}15\pi/8}}{2^{3/4}}\frac{e^{-\mathrm{i}\varphi/4}}{\varrho^{1/4}}
    + \frac{e^{\mathrm{i}\pi/8}}{2^{5/4}} \varrho^{1/4}
    e^{\mathrm{i}\varphi/4} + \mathrm{O}(\varrho^{3/4}) 
  \end{pmatrix}  \; .
\end{align*}
With a circle around the exceptional point ($\varphi = 0\dots 2\pi$) it
can be seen directly that $\bm{x}_1(2\pi) = - \bm{x}_2(0)$, and
$\bm{x}_2(2\pi) = \bm{x}_1(0)$, i.e., the permutation rule
\eqref{eq:sign_change} for a symmetric matrix is reproduced in this
picture.


\end{document}